\documentclass{article}

\usepackage{amsmath}
\usepackage{xspace}
\usepackage{color}
\usepackage{graphicx}
\usepackage{url}

\usepackage{caption} \usepackage{subcaption}

\newtheorem{theorem}{Theorem}
\newtheorem{proposition}{Proposition}
\newtheorem{definition}{Definition}
\newcommand{\putaway}[1]{}
\newcommand{\logic}[1]{{\normalfont\textsf{#1}}}
\newcommand{\tuple}[1]{\langle #1 \rangle}

\newcommand{\exto}[2]{|| #1 ||^{#2} }

\newcommand{\rconf}{\mathit{conf}}
\newcommand{\rdisc}{\mathit{disc}}
\newcommand{\rdabl}{\mathit{dabl}}

\newcommand{\evid}[1]{\textsc{conf}_{#1}}
\newcommand{\fals}[1]{\textsc{disc}_{#1}}
\newcommand{\dabl}[1]{\textsc{can}_{#1}}
\newcommand{\able}[1]{\textsc{able}_{#1}}
\newcommand{\task}[1]{\textsc{task}_{#1}}
\newcommand{\agree}[1]{\textsc{agree}_{#1}}
\newcommand{\limp}{\rightarrow}
\newcommand{\leqv}{\leftrightarrow}
\newcommand{\Agt}{{\normalfont\textsf{Agt}}\xspace}
\newcommand{\Prop}{{\normalfont\textsf{Prop}}\xspace}

\newcommand{\mana}[0]{{\sf m}}

\newcommand{\U}{\mathcal{U}}
\renewcommand{\S}{\mathcal{S}}
\newcommand{\G}{\mathcal{G}}
\newcommand{\F}{\mathcal{F}}
\renewcommand{\H}{\mathcal{H}}
\renewcommand{\P}{\mathcal{P}}
\newcommand{\W}{\mathcal{W}}

\newcommand{\para}[1]{\smallskip\noindent\textbf{#1.}}

\author{Nicolas Troquard\\
  {\small Faculty of Computer Science}\\
  {\small Free University of Bozen-Bolzano}\\ 
  {\small Piazza Domenicani, 3}\\
  {\small 39100 Bozen-Bolzano BZ, Italy}\\
  {\small {\tt ntroquard@unibz.it}}
}
\title{Tracking and managing deemed abilities}
\date{}

\begin{document}

\maketitle

\begin{abstract}
Information about the powers and abilities of acting entities is used to coordinate their actions in societies, either physical or digital.
Yet, the commonsensical meaning of an acting entity being deemed able to do something is still missing from the existing specification languages for the web or for multi-agent systems. We advance a general purpose abstract logical account of evidence-based ability. A basic model can be thought of as the ongoing trace of a multi-agent system. Every state records systemic confirmations and disconfirmations of whether an acting entity is able to bring about something. Qualitative inductive reasoning is then used in order to infer what acting entities are deemed able to bring about in the multi-agent system.
A temporalised modal language is used to talk about deemed ability, actual agency, and confirmation and disconfirmation of deemed ability.
What constitutes a confirmation and a disconfirmation is left to the modeller as in general it depends on the application at hand. So to illustrate the methodology we propose two extended examples, one in practical philosophy, the other in system engineering. We first use a logic of agency and ability to obtain a version of Mele's general practical abilities. Then, we look at the management of abilities in a supervised system.
\end{abstract}

\section{Preliminaries}
\label{sec:intro}

\noindent
In open and highly distributed environments like online services,
there are many unknown variables. Specifications are typically
provided by the services themselves, and their actual implementation
may depend on human intervention. Both are prone to error,
to misrepresentation \emph{bona fide}, and possibly, to deception. On a
service repository, services join, evolve, and leave. Think about the
sellers on Amazon Marketplace or AbeBooks: They start new
businesses, modify their business models, go bankrupt, on sick leave,
and on holiday.

\para{Scenario} 
{\em Imagine a rare media service repository. Now, a client of the service
repository wishes to acquire two original copies of \emph{Robinson
Crusoe}. Upon inspection, the server manager finds that services
$\sigma_1$, $\sigma_2$, and $\sigma_3$ each have one copy on
offer. This seems like a trivial enough choreography problem; a
typical choreography procedure could combine $\sigma_1$ and
$\sigma_2$ in order to obtain a complex service function which, one can
expect, consists in sending two copies of the book to the client. The
function is called. The outcome is that the client receives a
confirmation from $\sigma_1$ that the book has been dispatched and an
apologetic message from $\sigma_2$ indicating that it cannot fulfil
the request. $\sigma_2$ failed its part of the function. After
enquiry, the server manager discovers that $\sigma_2$ does not
actually have books in store, but instead acquires them from
$\sigma_3$ at discount price. The server manager also finds out that
$\sigma_3$ has ceased activity temporarily.}

\smallskip

We will come back to service science a few times for illustrative
purposes, but we are interested in the bigger picture. In settings
presenting the same challenges, how do we deal with the inherent
openness of such systems?  How do we evaluate the abilities of acting
entities?  Evaluation of power is subjective. It reflects the
information possessed and considered relevant by the system manager,
and his own logic of knowledge management. How do we maintain the
evaluation of abilities in accordance with the perceived changes in the
multi-agent system? \emph{Confirmation} and \emph{disconfirmation} are 
crucial notions in this connection.
Depending on the application, they can be the actual
exercise of some ability, or the omission thereof. The confirmation or disconfirmation of an ability can be a generalised
speech-act like a registration, a ban, the signing of a contract, or
the registration of a medical file of aptitude or
of invalidity. Remarkable work has been done in this direction in trust
assessment using numerical models, e.g.,
\cite{Wang:2010:ETM:1867713.1867715}. But at a more abstract level, we
are still compelled to understand and capture the logical aspects of
these mechanisms.

\medskip

Logics dedicated to the study of powers have been studied in Theoretical Computer Science and
Artificial Intelligence, with Alternating-time Temporal Logic (ATL)~\cite{alur02atl} being
probably the foremost representative. ATL and its numerous variants are
excellent formalisms to reason about concurrent systems where the
distributed components interact in a game theoretical
fashion. Judgements about power are derived from unambiguous models that
describe the \emph{a priori} knowledge that a designer has of an
interaction set-up. These models are \emph{concurrent game
  structures}. Of particular relevance to our illustrative scenario
may be their use in service composition~\cite{DeGiacomo:2010}. In
concurrent game structures, an acting entity is able to bring about that $\phi$ at some
moment if it possesses a pertinent action/strategy that would ensure
that $\phi$ when executed. Of course, a judgement about ability in this
setting does not say anything about the same power at earlier or later
times. But in less rigid societies of agents, judgements about ability
are often less specific and definitive than in game theoretical
models. Ability is more often merely discovered from experience: ``We
really do have an `idea' corresponding to the word `power'. [...] Hume
saw that recognition must be given to the essential part played by
`experience of the past' in our knowledge of the existence of
powers.''~\cite[p.~59]{ayers68}.
We acknowledge the
historical value of the statement, for we aim to propose a logical
framework that recognises it as fundamental. In this paper we focus
our attention on the ``can'' of power, and more precisely on the
commonsensical meaning of ``can'' as \emph{deemed ability} that
accounts for present, previous, and future \emph{confirmations} and
\emph{disconfirmations} of ability. The result is a work of logic 
that differs greatly from the existing formalisms like ATL.

\medskip
In a first part, we will present a general purpose formal framework to reason
faithfully about deemed ability. We will assume a linear flow of time
and the abstract modal notions of confirmation and disconfirmation. We will
present a model theory and a formal language to represent and express
specifically the following general principles of ``being deemed able''
(Section~\ref{sec:ebability}):
\begin{enumerate}
\item If the current situation provides the confirmation that the acting entity $G$ is able to bring about $\phi$ then $G$ is deemed able to bring about $\phi$;
\item If the current situation disconfirms that $G$ is able to bring
  about $\phi$ then $G$ is not deemed able to bring about $\phi$;
\item If an acting entity $G$ is deemed able to bring about $\phi$, it
  will continue to be deemed able unless and until we encounter a disconfirmation of this ability.
\item If an acting entity $G$ is not deemed able to bring about $\phi$,
  it remains so unless and until we encounter a confirmation for this ability;
\item If an acting entity $G$ is deemed able to bring about something, it
  is so because there is a confirmation of it now, or there has been
  a confirmation of it in then past and $G$ has been deemed able ever since.
\end{enumerate}

From the five principles above, we can observe that confirmations,
disconfirmation, and the flow of time are necessary and sufficient to
decide whether an acting entity is deemed able for something at some
instant. We consider the assumption of a linear flow of time to be
benign for all practical purpose. Then effectively, this means that
the task of specifying the powers of the acting entities of a system
can focus exclusively on the two \emph{empirical} notions:
confirmation and disconfirmation of ability. Hence, by adopting the simple
generic framework of ``being deemed able'' for a particular system
specification, one steers clear from metaphysical debate
about what constitutes an ability in the system at hand. 

Of course, there is no one-size-fits-all solution to tracking abilities
in multi-agent systems. What constitutes a confirmation and what constitutes a disconfirmation must depend on the application at hand.
So in a second part we will present more \emph{ad hoc} groundings of
confirmation and disconfirmation.  To illustrate the methodology we propose
two extended examples, one in practical philosophy, the other in
system engineering. First we use a logic of agency and ability to
obtain a version of Mele's general practical abilities
(Section~\ref{sec:elgtemp}). Then, we look at the management of abilities
in a supervised system (Section~\ref{sec:pract}).

\section{Being deemed able: the core logic}
\label{sec:ebability}

We suppose a finite supply of \emph{agents} collected in a set \Agt.
A \emph{group} is any subset of \Agt. We call an \emph{acting
  entity} any agent or group.
We also suppose an infinite supply of propositional variables collected in a set \Prop. The sets $\Agt$ and $\Prop$ are fixed throughout the paper.

We first define the \emph{static} core logic. Then, we give a
temporalisation of it. Then, we extend it to obtain the core logic of
being deemed able.

\subsection{The static core logic}
In the following we will use three linguistic constructs that are at
the core of the logic of being deemed able. $\dabl{G}\phi$ reads
``acting entity $G$ is deemed able to bring about that
$\phi$''. $\evid{G}\phi$ reads ``the situation confirms that
acting entity $G$ is able to bring about that $\phi$''.  $\fals{G}\phi$
reads ``the situation disconfirms that acting entity $G$ is able to bring it
about that $\phi$''. These readings will also often be rephrased throughout the paper into contextually more appropriate wordings.


\medskip

To talk about the static facts of being deemed able, we extend the
language of propositional logic with the three previous
modalities. Formally, we obtain the language $L_\logic{sc}$
(where $p \in \Prop$ and $G \subseteq \Agt$) with the following
grammar in Backus-Naur form:
\[
\phi ::= p  \ \mid\  \lnot \phi  \ \mid\  \phi \land \phi  \ \mid\  \dabl{G} \phi \ \mid\  \evid{G} \phi  \ \mid\  \fals{G}\phi
\]
Throughout the paper, we use $\phi \lor \psi$ as a notational variant of $\lnot (\lnot \phi \land \lnot \psi)$ and
$\phi \limp \psi$ as a notational variant of $\lnot \phi \lor \psi$.

A formula in $L_\logic{sc}$ can contain arbitrary nestings of modalities. As for any expressive enough language, some grammatically correct sentences could be gibberish, or can be difficult to interpret in plain English: e.g., $\evid{G}\fals{G}\dabl{G} p$. On the other hand, some other combinations can be useful: e.g., $\dabl{G_1}\evid{G_2}p$ characterises a situation where the group~$G_1$ is deemed able to bring about a situation that is confirmation of the fact that the group~$G_2$ is able to bring about that $p$ holds.

\begin{table}
\begin{center}
\framebox{\small
\begin{tabular}{ll}
 {\bf [prop]} & an axiomatisation of classical propositional logic\\
 {\bf [sc1]} & $\vdash_\logic{sc} \evid{G}\phi \limp \dabl{G}\phi$\\
 {\bf [sc2]} & $\vdash_\logic{sc} \fals{G}\phi \limp \lnot \dabl{G}\phi$\\
 {\bf [scr1]} & if $\vdash_\logic{sc} \phi \leqv \psi$ then $\vdash_\logic{sc} \dabl{G}\phi \leqv \dabl{G}\psi$\\
 {\bf [scr2]} & if $\vdash_\logic{sc} \phi \leqv \psi$ then $\vdash_\logic{sc} \evid{G}\phi \leqv \evid{G}\psi$\\
 {\bf [scr3]} & if $\vdash_\logic{sc} \phi \leqv \psi$ then $\vdash_\logic{sc} \fals{G}\phi \leqv \fals{G}\psi$\\
\end{tabular}
}
\end{center}
\caption{$\vdash_\logic{sc}$\label{tab:scp}}
\end{table}

\medskip
The static core logic $\logic{sc}$ is the minimal set of formulas
closed under $\vdash_\logic{sc}$, presented in Table~\ref{tab:scp}.
Note that confirmation and disconfirmation with respect to the same ability are mutually
exclusive. By axiom~sc1, axiom~sc2, and classical logic, we have
$\vdash_\logic{sc} \evid{G}\phi \land \fals{G}\phi \limp \dabl{G}{\phi} \land \lnot \dabl{G}{\phi}$
and thus $\vdash_\logic{sc} \evid{G}\phi \land \fals{G}\phi
\limp \bot$.

For many practical purposes, we could close the confirmation and
disconfirmation operators under conjunctions. But we refrain from doing so,
to allow more modelling versatility. 


\medskip

We can provide a model theory for $\vdash_\logic{sc}$ with very simple
structures (we note $\mathcal{P}(W)$ the set of subsets of a set $W$):
\begin{definition}\label{def:scmodel}
An $\logic{sc}$-model is a tuple $M = \tuple{W,\rdabl,\rconf,\rdisc,V}$,
where for every $w \in W$ and $G \subseteq \Agt$,
$\rdabl(w)(G) \subseteq \mathcal{P}(W)$, $\rconf(w)(G) \subseteq
\mathcal{P}(W)$, $\rdisc(w)(G) \subseteq \mathcal{P}(W)$, and
$V(w) \subseteq \Prop$. In addition, it satisfies the following
constraints:
\begin{enumerate}
\item if $X \in \rconf(w)(G)$ then $X \in \rdabl(w)(G)$
\item if $X \in \rdisc(w)(G)$ then $X \not\in \rdabl(w)(G)$
\end{enumerate}
\end{definition}
We define the interpretation $\models_\logic{sc}$ of the language
$L_\logic{sc}$ in an $\logic{sc}$-model $M = \tuple{W,\rdabl,\rconf,\rdisc,V}$ as
follows:
\begin{itemize}
\item $M, w \models_\logic{sc} p$  iff  $p \in V(w)$
\item $M, w \models_\logic{sc} \lnot \phi$  iff  not $M, w \models_\logic{sc}  \phi$
\item $M, w \models_\logic{sc} \phi \land \psi$  iff  $M, w \models_\logic{sc} \phi$ and $M, w
  \models_\logic{sc}\psi$
\item $M, w \models_\logic{sc} \dabl{G}\phi$  iff  $\exto{\phi}{M} \in \rdabl(w)(G)$
\item $M, w \models_\logic{sc} \evid{G}\phi$  iff  $\exto{\phi}{M} \in \rconf(w)(G)$
\item $M, w \models_\logic{sc} \fals{G}\phi$  iff  $\exto{\phi}{M} \in \rdisc(w)(G)$
\end{itemize}
where $\exto{\phi}{M} = \{ w \mid M,w \models_\logic{sc} \phi\}$.

\medskip

It is routine to prove that the logic $\logic{sc}$ is sound and
complete wrt.\ the class of $\logic{sc}$-models~\cite{chellas80}.
\begin{proposition}\label{prop:complsc}
Let $\phi \in L_\logic{sc}$. Then, $\vdash_\logic{sc} \phi$ iff $\models_\logic{sc} \phi$.
\end{proposition}

\medskip

Effectively, the two constraints in Definition~\ref{def:scmodel} correspond to imposing the static principle linking a confirmation in a world to a deemed ability in that world
\begin{center}
  \begin{picture}(0,0)%
\includegraphics{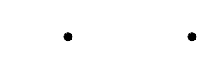}%
\end{picture}%
\setlength{\unitlength}{2901sp}%
\begingroup\makeatletter\ifx\SetFigFont\undefined%
\gdef\SetFigFont#1#2#3#4#5{%
  \reset@font\fontsize{#1}{#2pt}%
  \fontfamily{#3}\fontseries{#4}\fontshape{#5}%
  \selectfont}%
\fi\endgroup%
\begin{picture}(2143,664)(386,-3041)
\put(586,-2761){\makebox(0,0)[rb]{\smash{{\SetFigFont{8}{9.6}{\rmdefault}{\mddefault}{\updefault}{\color[rgb]{0,0,0}if}%
}}}}
\put(1126,-2536){\makebox(0,0)[b]{\smash{{\SetFigFont{8}{9.6}{\rmdefault}{\mddefault}{\updefault}{\color[rgb]{0,0,0}$\evid{G}\phi$}%
}}}}
\put(2476,-2536){\makebox(0,0)[b]{\smash{{\SetFigFont{8}{9.6}{\rmdefault}{\mddefault}{\updefault}{\color[rgb]{0,0,0}$\dabl{G}\phi$}%
}}}}
\put(1666,-2806){\makebox(0,0)[lb]{\smash{{\SetFigFont{8}{9.6}{\rmdefault}{\mddefault}{\updefault}{\color[rgb]{0,0,0}then}%
}}}}
\put(1126,-2986){\makebox(0,0)[b]{\smash{{\SetFigFont{8}{9.6}{\rmdefault}{\mddefault}{\updefault}{\color[rgb]{0,0,0}$w$}%
}}}}
\put(2476,-2986){\makebox(0,0)[b]{\smash{{\SetFigFont{8}{9.6}{\rmdefault}{\mddefault}{\updefault}{\color[rgb]{0,0,0}$w$}%
}}}}
\end{picture}%

\end{center}
and the static principle linking a disconfirmation in a world to an absence of deemed ability in that world.
\begin{center}
  \begin{picture}(0,0)%
\includegraphics{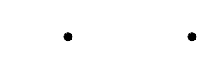}%
\end{picture}%
\setlength{\unitlength}{2901sp}%
\begingroup\makeatletter\ifx\SetFigFont\undefined%
\gdef\SetFigFont#1#2#3#4#5{%
  \reset@font\fontsize{#1}{#2pt}%
  \fontfamily{#3}\fontseries{#4}\fontshape{#5}%
  \selectfont}%
\fi\endgroup%
\begin{picture}(2143,664)(386,-3041)
\put(586,-2761){\makebox(0,0)[rb]{\smash{{\SetFigFont{8}{9.6}{\rmdefault}{\mddefault}{\updefault}{\color[rgb]{0,0,0}if}%
}}}}
\put(1126,-2536){\makebox(0,0)[b]{\smash{{\SetFigFont{8}{9.6}{\rmdefault}{\mddefault}{\updefault}{\color[rgb]{0,0,0}$\fals{G}\phi$}%
}}}}
\put(2476,-2986){\makebox(0,0)[b]{\smash{{\SetFigFont{8}{9.6}{\rmdefault}{\mddefault}{\updefault}{\color[rgb]{0,0,0}$w$}%
}}}}
\put(2476,-2536){\makebox(0,0)[b]{\smash{{\SetFigFont{8}{9.6}{\rmdefault}{\mddefault}{\updefault}{\color[rgb]{0,0,0}$\lnot\dabl{G}\phi$}%
}}}}
\put(1666,-2806){\makebox(0,0)[lb]{\smash{{\SetFigFont{8}{9.6}{\rmdefault}{\mddefault}{\updefault}{\color[rgb]{0,0,0}then}%
}}}}
\put(1126,-2986){\makebox(0,0)[b]{\smash{{\SetFigFont{8}{9.6}{\rmdefault}{\mddefault}{\updefault}{\color[rgb]{0,0,0}$w$}%
}}}}
\end{picture}%
\end{center}

The formula $\lnot \evid{G}\phi \land \dabl{G}\phi$ is
consistent in the static core logic.  The static core theory is thus
insufficient in that it does not explain why a group is deemed able.
When it holds, the theory does not say how this deemed ability came to
exist, and how it has been maintained \emph{before} and until now. The
theory does not say what this ability becomes \emph{after} an acting
entity is deemed able to do something.  If it is not deemed able to do
something, our theory does not say what it takes for it to be deemed
able at a later stage.

As a matter of fact, we do not even have the means to talk about
before and after in the static logic. We remedy this in the
remainder of this section.

\subsection{Linear until-since logic}

To talk and reason about temporal properties, we use a linear until-since logic based on the following language $L_\logic{time}$:
\[
\begin{array}{lcccccccccl}
\phi & ::= & p & \mid & \lnot \phi & \mid & \phi \land \phi & \mid & \phi \U \phi & \mid & \phi \S \phi
\end{array}
\]
The formula $\phi \U \psi$ reads that the property $\phi$ holds at
least until $\psi$ is true. 
With it, one can define the usual ``eventually/future'' operator $\F\phi = \top \U \phi$ and  
the ``always/globally'' operator, $\G\phi = \lnot \F \lnot \phi$.
We can also define the ``weak until'' as $\phi \W\psi = (\phi \U \psi) \lor \G\phi$, which will be particularly useful in this paper. 
The ``since'' operator $\S$ is used to talk about the past. The formula
$\phi \S \psi$ reads that the proposition $\phi$ has been holding ever
since $\psi$ was true. 
With it, one can define the ``has always been''
operator, $\H\phi = \lnot \P \lnot \phi$ and 
``has been in the past'' operator $\P\phi = \top \S \phi$.

\begin{definition}
A \emph{flow of time} is a tuple $\tuple{T,<}$ where $T$ is a nonempty
set of instants and $<$ is a linear
order over $T$. A \emph{model of time} is a tuple $F
= \tuple{T,<,g}$, where $\tuple{T,<}$ is a flow of time and $g$ is a
valuation function such that $g(t) \subseteq \Prop$ for all $t \in T$.
\end{definition}



\medskip

The interpretation $\models_\logic{time}$ of the language
$L_\logic{time}$ in a model of time $M = {\tuple{T,<,g}}$ is defined as
follows~\cite{kamp71}:
\begin{itemize}
\item $M,t \models_\logic{time} p$ iff $p \in g(t)$, when $p \in \Prop$

\item $M, t \models_\logic{time} \phi \U \psi$ iff there is an $s \in T$ with $t <
  s$ and $M,s \models_\logic{time} \psi$ and for every $u \in T$, if $t < u < s$
  then $M,u \models_\logic{time} \phi$

\item $M, t \models_\logic{time} \phi \S \psi$ iff there is an $s \in T$ with $s <
  t$ and $M,s \models_\logic{time} \psi$ and for every $u \in T$, if $s < u < t$
  then $M,u \models_\logic{time} \phi$
\end{itemize}

\medskip

\begin{table}
\begin{center}
\framebox{{\small
 \begin{tabular}{ll}
 {\bf [prop]} & an axiomatisation of classical propositional logic\\
 {\bf [ltl1]} & $\vdash_\logic{time} \G(p \limp q) \land (r \U p) \limp (r \U q)$\\
 {\bf [ltl2]}& $\vdash_\logic{time} \H(p \limp q) \land (r \S p) \limp (r \S q)$\\
 {\bf [ltl3]}& $\vdash_\logic{time} \G(p \limp q) \land (p \U r) \limp (q \U r)$\\
 {\bf [ltl4]}& $\vdash_\logic{time} \H(p \limp q) \land (p \S r) \limp (q \S r)$\\
 {\bf [ltl5]}& $\vdash_\logic{time} p \land (r \U q) \limp (r \U (q \land (r \S p)))$\\
 {\bf [ltl6]}& $\vdash_\logic{time} p \land (r \S q) \limp (r \S (q \land (r \U p)))$\\
 {\bf [ltl7]}& $\vdash_\logic{time} q \U p \limp (q \land (q \U p)) \U p$\\
 {\bf [ltl8]}& $\vdash_\logic{time} q \S p \limp (q \land (q \S p)) \U p$\\
 {\bf [ltl9]}& $\vdash_\logic{time} (q \U (q \land (q \U p))) \limp (q \U p)$\\
 {\bf [ltl10]}& $\vdash_\logic{time} (q \S (q \land (q \S p))) \limp (q \S p)$\\
 {\bf [ltl11]}&  $\vdash_\logic{time} ((q \U p) \land (s \U r)) \limp (((q \land s) \U (p\land r)) \lor$\\&\hfill$ 
                                            ((q \land s) \U (p\land s)) \lor
                                            ((q \land s) \U (q\land r)))$\\
 {\bf [ltl12]}&  $\vdash_\logic{time} ((q \S p) \land (s \S r)) \limp (((q \land s) \S (p\land r)) \lor$\\&\hfill$ 
                                            ((q \land s) \S (p\land s)) \lor
                                            ((q \land s) \S (q\land r)))$\\
 {\bf [ltlr1]}& if $\vdash_\logic{time} \phi$ then $\vdash_\logic{time} \G\phi$\\
 {\bf [ltlr2]}& if $\vdash_\logic{time} \phi$ then $\vdash_\logic{time} \H\phi$\\
\end{tabular}}
}
\end{center}
\caption{$\vdash_\logic{time}$\label{tab:timep}}
\end{table}

The tense logic $\logic{time}$ is the minimal set of formulas closed
under $\vdash_\logic{time}$, presented on Table~\ref{tab:timep}. 
Axioms~ltl11 and ltl12 ensure
that the time is linear. Axioms~ltl7, ltl8, and ltl9 ensure that time is
transitive. Xu showed in~\cite{xu88:_some_u_s_tense_logic} that ltl10
is in fact redundant.
An axiomatisation of the tense logic $\logic{time}$ is presented
in~\cite{xu88:_some_u_s_tense_logic}.  
The following theorem is due to Burgess~\cite{burgess82:_axiom}, and
Xu~\cite{xu88:_some_u_s_tense_logic}.
\begin{theorem}\label{th:compltime}
Let $\phi \in L_\logic{time}$. Then, $\vdash_\logic{time} \phi$ iff $\models_\logic{time} \phi$.
\end{theorem}

\subsection{Temporalisation} We now present the temporalisation of the static core logic of being deemed able with the until-since logic.

\begin{definition}
A formula $\phi \in L_\logic{sc}$ is a \emph{Boolean combination} iff
it is built up from other formulas by means of the Boolean connectives
$\land$ and $\lnot$ or any other connectives defined in terms of
those.  A formula $\alpha \in L_\logic{sc}$ is a \emph{monolithic
  formula} iff it is not a Boolean combination.
\end{definition}
Examples of Boolean combinations are $\lnot p$, or $\dabl{G_1}p \limp \lnot \fals{G_2}q$.
Examples of monolithic formulas are: $p$, $\fals{G}(q \limp p)$,
or $\dabl{G_1}(\evid{G_1}(p \lor q) \land \lnot\fals{G_2}q)$.

Following~\cite{finger92jolli}, we temporalise $\logic{sc}$ with
$\logic{time}$, and obtain the logic system
$\logic{time}(\logic{sc})$. We need to define its language, proof
theory, models, and model theory.

The language $L_{\logic{time}(\logic{sc})}$ is defined as follows:
\[
\begin{array}{lcccccccccl}
\phi & ::= & \alpha & \mid & \lnot \phi & \mid & \phi \land \phi & \mid & \phi \U \phi & \mid & \phi \S \phi
\end{array}
\]
where $\alpha$ is a monolithic formula of $L_\logic{sc}$.

The proof theory $\vdash_{\logic{time}(\logic{sc})}$ consists of:
\begin{itemize}
\item all the principles of $\logic{time}$, 
and 
\item if $\vdash_\logic{sc} \phi$ then
  $\vdash_{\logic{time}(\logic{sc})} \phi$, when $\phi \in
  L_\logic{sc}$
\end{itemize}

We define the models of $\logic{time}(\logic{sc})$.
\begin{definition}\label{def:timescmodel}
A model of $\logic{time}(\logic{sc})$ is a tuple
$M_{\logic{time}(\logic{sc})} = \tuple{T,<,g}$ where $\tuple{T,<}$ is
a flow of time, and $g$ a function that maps every member
of $T$ into a pointed model $(M, w)$, with $M =
\tuple{W,\rdabl,\rconf,\rdisc,V}$ an $\logic{sc}$-model and $w \in W$.
\end{definition}
The interpretation of $L_{\logic{time}(\logic{sc})}$ in a model $M_{\logic{time}(\logic{sc})} = \tuple{T,<,g}$ of $\logic{time}(\logic{sc})$ is simply:
\begin{itemize}
\item $M_{\logic{time}(\logic{sc})},t \models \alpha$ iff $g(t)
  \models_\logic{sc} \alpha$ 
\end{itemize}
when $\alpha$ is a monolithic formula, and analogous to $\models_\logic{time}$ otherwise, while the truth value of the temporal operators is as before.
\begin{figure}
\begin{center}
  \begin{picture}(0,0)%
\includegraphics{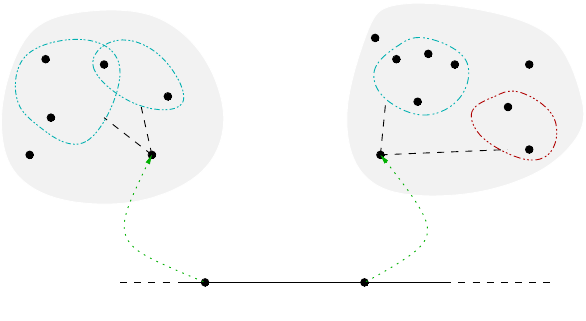}%
\end{picture}%
\setlength{\unitlength}{3729sp}%
\begingroup\makeatletter\ifx\SetFigFont\undefined%
\gdef\SetFigFont#1#2#3#4#5{%
  \reset@font\fontsize{#1}{#2pt}%
  \fontfamily{#3}\fontseries{#4}\fontshape{#5}%
  \selectfont}%
\fi\endgroup%
\begin{picture}(4955,2701)(1190,-3761)
\put(2926,-3706){\makebox(0,0)[b]{\smash{{\SetFigFont{9}{10.8}{\rmdefault}{\mddefault}{\updefault}{\color[rgb]{0,0,0}$t$}%
}}}}
\put(4276,-3706){\makebox(0,0)[b]{\smash{{\SetFigFont{9}{10.8}{\rmdefault}{\mddefault}{\updefault}{\color[rgb]{0,0,0}$t'$}%
}}}}
\put(2161,-3031){\makebox(0,0)[rb]{\smash{{\SetFigFont{11}{13.2}{\rmdefault}{\mddefault}{\updefault}{\color[rgb]{0,0,0}$g$}%
}}}}
\put(4951,-3031){\makebox(0,0)[lb]{\smash{{\SetFigFont{11}{13.2}{\rmdefault}{\mddefault}{\updefault}{\color[rgb]{0,0,0}$g$}%
}}}}
\put(4276,-2986){\makebox(0,0)[b]{\smash{{\SetFigFont{9}{10.8}{\rmdefault}{\mddefault}{\updefault}{\color[rgb]{0,0,0}$\dabl{G}\psi$}%
}}}}
\put(4276,-3211){\makebox(0,0)[b]{\smash{{\SetFigFont{9}{10.8}{\rmdefault}{\mddefault}{\updefault}{\color[rgb]{0,0,0}$\fals{G}\phi$}%
}}}}
\put(4096,-2671){\makebox(0,0)[rb]{\smash{{\SetFigFont{11}{13.2}{\rmdefault}{\mddefault}{\updefault}{\color[rgb]{0,0,0}$M'_s$}%
}}}}
\put(2566,-2491){\makebox(0,0)[lb]{\smash{{\SetFigFont{11}{13.2}{\rmdefault}{\mddefault}{\updefault}{\color[rgb]{0,0,0}$w$}%
}}}}
\put(2521,-2131){\makebox(0,0)[lb]{\smash{{\SetFigFont{7}{8.4}{\rmdefault}{\mddefault}{\updefault}{\color[rgb]{0,0,0}$\rdabl$}%
}}}}
\put(4951,-2491){\makebox(0,0)[lb]{\smash{{\SetFigFont{7}{8.4}{\rmdefault}{\mddefault}{\updefault}{\color[rgb]{0,0,0}$\rdisc$}%
}}}}
\put(4501,-2176){\makebox(0,0)[lb]{\smash{{\SetFigFont{7}{8.4}{\rmdefault}{\mddefault}{\updefault}{\color[rgb]{0,0,0}$\rdabl$}%
}}}}
\put(4726,-1771){\makebox(0,0)[b]{\smash{{\SetFigFont{6}{7.2}{\rmdefault}{\mddefault}{\updefault}{\color[rgb]{0,0,0}$\exto{\psi}{M'_s}$}%
}}}}
\put(2476,-1771){\makebox(0,0)[b]{\smash{{\SetFigFont{6}{7.2}{\rmdefault}{\mddefault}{\updefault}{\color[rgb]{0,0,0}$\exto{\psi}{M_s}$}%
}}}}
\put(2926,-3301){\makebox(0,0)[b]{\smash{{\SetFigFont{9}{10.8}{\rmdefault}{\mddefault}{\updefault}{\color[rgb]{0,0,0}$\dabl{G}\phi$}%
}}}}
\put(2926,-3121){\makebox(0,0)[b]{\smash{{\SetFigFont{9}{10.8}{\rmdefault}{\mddefault}{\updefault}{\color[rgb]{0,0,0}$\dabl{G}\psi$}%
}}}}
\put(2926,-2941){\makebox(0,0)[b]{\smash{{\SetFigFont{9}{10.8}{\rmdefault}{\mddefault}{\updefault}{\color[rgb]{0,0,0}$\evid{G}\psi$}%
}}}}
\put(2521,-2266){\makebox(0,0)[lb]{\smash{{\SetFigFont{7}{8.4}{\rmdefault}{\mddefault}{\updefault}{\color[rgb]{0,0,0}$\rconf$}%
}}}}
\put(2296,-2311){\makebox(0,0)[rb]{\smash{{\SetFigFont{7}{8.4}{\rmdefault}{\mddefault}{\updefault}{\color[rgb]{0,0,0}$\rdabl$}%
}}}}
\put(5581,-2131){\makebox(0,0)[b]{\smash{{\SetFigFont{6}{7.2}{\rmdefault}{\mddefault}{\updefault}{\color[rgb]{0,0,0}$\exto{\phi}{M'_s}$}%
}}}}
\put(1711,-1861){\makebox(0,0)[b]{\smash{{\SetFigFont{6}{7.2}{\rmdefault}{\mddefault}{\updefault}{\color[rgb]{0,0,0}$\exto{\phi}{M_s}$}%
}}}}
\put(4366,-2311){\makebox(0,0)[rb]{\smash{{\SetFigFont{11}{13.2}{\rmdefault}{\mddefault}{\updefault}{\color[rgb]{0,0,0}$w'$}%
}}}}
\put(2926,-2626){\makebox(0,0)[lb]{\smash{{\SetFigFont{11}{13.2}{\rmdefault}{\mddefault}{\updefault}{\color[rgb]{0,0,0}$M_s$}%
}}}}
\end{picture}%
\end{center}
\caption{\label{fig:tempo} Illustration of temporalisation.}
\end{figure}
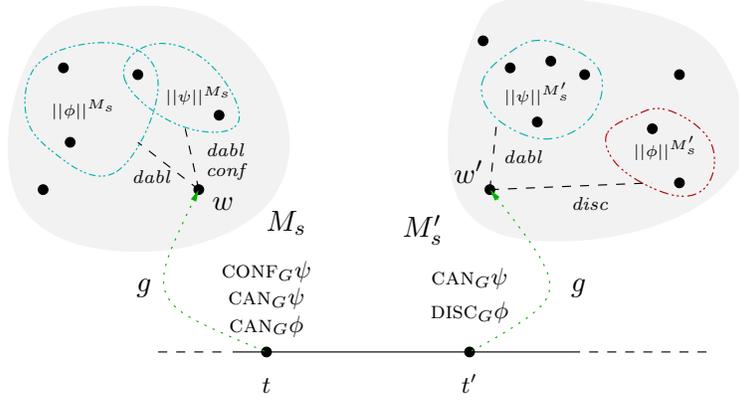

\medskip
Figure~\ref{fig:tempo} illustrates the temporalisation of the static core logic of being deemed able with the until-since logic. At time~$t$, the corresponding 
pointed $\logic{sc}$-model is $g(t) = (M_s,w)$. The figure represents the fact that $\exto{\phi}{M_s} \in \rdabl(w)(G)$.
It means that $M_s, w \models_\logic{sc} \dabl{G}{\phi}$, and since $\dabl{G}{\phi}$ is a monolithic formula, we also have that $M,t \models_{\logic{time}(\logic{sc})} \dabl{G}{\phi}$. For similar reasons, we also have that
$M,t \models_{\logic{time}(\logic{sc})} \evid{G}{\psi}$, and $M,t \models_{\logic{time}(\logic{sc})} \dabl{G}{\psi}$.
At time~$t'$, the corresponding pointed $\logic{sc}$-model is $g(t') = (M'_s,w')$. The figure represents the fact that $\exto{\phi}{M'_s} \in \rdisc(w')(G)$, and that $\exto{\psi}{M'_s} \in \rdabl(w')(G)$. We have that $M,t \models_{\logic{time}(\logic{sc})} \dabl{G}{\psi}$ and $M,t \models_{\logic{time}(\logic{sc})} \fals{G}{\phi}$.

Hence, when $t \not = t'$, the pointed models $g(t)$ and $g(t')$ can be different (as exemplified in Figure~\ref{fig:tempo}), but not necessarily so. In them, abilities, confirmations, and disconfirmations are possibly given different truth values, but not necessarily so. The models of $\logic{time}(\logic{sc})$ do not impose any constraints. To the contrary, the models of deemed ability that we introduce in Section~\ref{sec:lbda} will be models of $\logic{time}(\logic{sc})$ satisfying specific temporal constraints about abilities, confirmations, and disconfirmations.

\medskip
The following result is an immediate consequence of
Proposition~\ref{prop:complsc}, Theorem~\ref{th:compltime} and the
general completeness theorem of temporalisations due to Finger and
Gabbay~\cite[Th.~2.3]{finger92jolli}:
\begin{proposition}\label{th:thcompl-timesc}
If $\phi \in L_{\logic{time}(\logic{sc})}$,  $\vdash_{\logic{time}(\logic{sc})} \phi$ iff $\models_{\logic{time}(\logic{sc})} \phi$.
\end{proposition}

\subsection{The core logic of being deemed able}\label{sec:lbda}
The static core logic of being deemed able already captures the first
two principles listed in Section~\ref{sec:intro}. We now extend
$\logic{time}(\logic{sc})$ to capture the temporal principles. Let us
call $\logic{lbda}$ the core logic of being deemed able. We assume:
\begin{equation*}
  \text{if } \vdash_{\logic{time}(\logic{sc})} \phi \text{ then } \vdash_{\logic{lbda}} \phi
  \tag{\textbf{lbdar1}}
\end{equation*}
We can now formalise the last three principles of being deemed able
that we listed in the introduction.


\para{The dynamic role of disconfirmation}\label{sec:falsification}
When an acting entity is deemed able to do something, one can maintain this
perceived ability \emph{until} some further evidence \emph{disconfirms}
it.
This suggests the following axiom:
\begin{equation*}
\vdash_{\logic{lbda}} \dabl{G} \phi
\limp (\dabl{G} \phi) \W (\fals{G}\phi)
  \tag{\textbf{lbda1}}
\end{equation*}
In words, if $G$ is deemed able to do $\phi$, it is deemed able until a
disconfirming situation occurs. When this situation occurs, we shall have
$\lnot\dabl{G}\phi$ by axiom~sc2. Note that we use the weak version of
the ``until'' operator $\W$. This is to capture the fact that
$\dabl{G}\phi$ might never actually be false in the future, which by
axiom~sc2, would mean that a disconfirmation never actually occurs in
the future. An existing deemed ability which is never disconfirmed is after all
the best of deemed abilities. Using $\U$, a disconfirmation would necessarily
have to occur in the future in order for an acting entity to be able to do
something. This would be counter-intuitive.

\para{The dynamic role of confirmation}\label{sec:showevidence}
We have just explained how an ability is maintained once it is deemed
to exist. If an acting entity is \emph{not} deemed able to bring about
something, how do we maintain this inability?  We adopt the following
principle, that is symmetrical to~lbda1.
\begin{equation*}
\vdash_{\logic{lbda}} \lnot \dabl{G}
\phi \limp (\lnot \dabl{G} \phi) \W (\evid{G} \phi)
  \tag{\textbf{lbda2}}
\end{equation*}

In words, if $G$ is not deemed able to do $\phi$, then it will not
be deemed able until a situation is reached that shows evidence of its deemed ability. Note the use of a weak ``until'' again. If this situation
showing confirmation of deemed ability is ever reached, we shall have $\dabl{G}\phi$ by axiom~sc1.

It remains to address what must be the past chronicle of an existing
ability. An entity $G$ is deemed able to do $\phi$ only if it has been so
ever since the occurrence of a situation showing confirmation for it.
\begin{equation*}
 \vdash_{\logic{lbda}} \dabl{G} \phi \limp (\evid{G}\phi) \lor
 ((\dabl{G} \phi) \S (\evid{G} \phi))
 \tag{\textbf{lbda3}}
\end{equation*}

Notice that we use here a standard ``since'' temporal operator $\S$, as
opposed to the weak ``until'' $\U$ in the principles~lbda1 and~lbda2. By
doing so, we commit our theory to the assumption that the existence of
a deemed ability has to be grounded on confirmation. 
It rules out the possibility that it is \emph{a priori} for some acting entity to be deemed able to bring about a contingent state of affairs.
The first disjunction on the right hand of the
implication captures the possibility that the current situation is one
showing the pertinent evidence.

\para{The models of deemed ability}
We can now constrain the models of the logic $\logic{time}(\logic{sc})$ 
so as to satisfy the principles for
the dynamic role of confirmation and of disconfirmation of ability.
We define the models of deemed ability.
\begin{definition}\label{def:model-da}
A \emph{model of deemed ability} is a model $M = \tuple{T,<,g}$ of
$\logic{time}(\logic{sc})$, that satisfies the following constraints
C1, C2 and C3:\\

\noindent\textbf{C1}~  if $M,t \models_{\logic{time}(\logic{sc})} \dabl{G}\phi$ then (i)~there is $t < t'$, such that  $M,t' \models_{\logic{time}(\logic{sc})} \fals{G}\phi$ and $M,t'' \models_{\logic{time}(\logic{sc})} \dabl{G}\phi$ for all $t < t'' < t'$, or (ii)~for every $t < t'$ we have  $M,t' \models_{\logic{time}(\logic{sc})} \dabl{G}\phi$.

\begin{center}
  \begin{picture}(0,0)%
\includegraphics{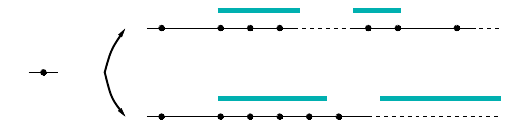}%
\end{picture}%
\setlength{\unitlength}{2072sp}%
\begingroup\makeatletter\ifx\SetFigFont\undefined%
\gdef\SetFigFont#1#2#3#4#5{%
  \reset@font\fontsize{#1}{#2pt}%
  \fontfamily{#3}\fontseries{#4}\fontshape{#5}%
  \selectfont}%
\fi\endgroup%
\begin{picture}(7703,2035)(463,-3725)
\put(2926,-2356){\makebox(0,0)[b]{\smash{{\SetFigFont{6}{7.2}{\rmdefault}{\mddefault}{\updefault}{\color[rgb]{0,0,0}$t$}%
}}}}
\put(7426,-2356){\makebox(0,0)[b]{\smash{{\SetFigFont{6}{7.2}{\rmdefault}{\mddefault}{\updefault}{\color[rgb]{0,0,0}$t'$}%
}}}}
\put(4726,-2356){\makebox(0,0)[b]{\smash{{\SetFigFont{6}{7.2}{\rmdefault}{\mddefault}{\updefault}{\color[rgb]{0,0,0}$t''$}%
}}}}
\put(2926,-1861){\makebox(0,0)[b]{\smash{{\SetFigFont{6}{7.2}{\rmdefault}{\mddefault}{\updefault}{\color[rgb]{0,0,0}$\dabl{G}\phi$}%
}}}}
\put(7426,-1861){\makebox(0,0)[b]{\smash{{\SetFigFont{6}{7.2}{\rmdefault}{\mddefault}{\updefault}{\color[rgb]{0,0,0}$\fals{G}\phi$}%
}}}}
\put(5446,-1861){\makebox(0,0)[b]{\smash{{\SetFigFont{6}{7.2}{\rmdefault}{\mddefault}{\updefault}{\color[rgb]{0,0,0}$\dabl{G}\phi$}%
}}}}
\put(2926,-3661){\makebox(0,0)[b]{\smash{{\SetFigFont{6}{7.2}{\rmdefault}{\mddefault}{\updefault}{\color[rgb]{0,0,0}$t$}%
}}}}
\put(2926,-3211){\makebox(0,0)[b]{\smash{{\SetFigFont{6}{7.2}{\rmdefault}{\mddefault}{\updefault}{\color[rgb]{0,0,0}$\dabl{G}\phi$}%
}}}}
\put(5851,-3211){\makebox(0,0)[b]{\smash{{\SetFigFont{6}{7.2}{\rmdefault}{\mddefault}{\updefault}{\color[rgb]{0,0,0}$\dabl{G}\phi$}%
}}}}
\put(1126,-2986){\makebox(0,0)[b]{\smash{{\SetFigFont{6}{7.2}{\rmdefault}{\mddefault}{\updefault}{\color[rgb]{0,0,0}$t$}%
}}}}
\put(586,-2761){\makebox(0,0)[rb]{\smash{{\SetFigFont{6}{7.2}{\rmdefault}{\mddefault}{\updefault}{\color[rgb]{0,0,0}if}%
}}}}
\put(1576,-2671){\makebox(0,0)[lb]{\smash{{\SetFigFont{6}{7.2}{\rmdefault}{\mddefault}{\updefault}{\color[rgb]{0,0,0}then}%
}}}}
\put(1126,-2536){\makebox(0,0)[b]{\smash{{\SetFigFont{6}{7.2}{\rmdefault}{\mddefault}{\updefault}{\color[rgb]{0,0,0}$\dabl{G}\phi$}%
}}}}
\put(1531,-2896){\makebox(0,0)[lb]{\smash{{\SetFigFont{6}{7.2}{\rmdefault}{\mddefault}{\updefault}{\color[rgb]{0,0,0}(or)}%
}}}}
\end{picture}%
\end{center}

\noindent\textbf{C2}~  if $M,t \models_{\logic{time}(\logic{sc})} \lnot\dabl{G}\phi$ then (i)~there is $t < t'$, such that  $M,t' \models_{\logic{time}(\logic{sc})} \evid{G}\phi$ and $M,t'' \models_{\logic{time}(\logic{sc})} \lnot\dabl{G}\phi$ for all $t < t'' < t'$, or (ii)~for every $t < t'$ we have  $M,t' \models_{\logic{time}(\logic{sc})} \lnot\dabl{G}\phi$.

\begin{center}
  \begin{picture}(0,0)%
\includegraphics{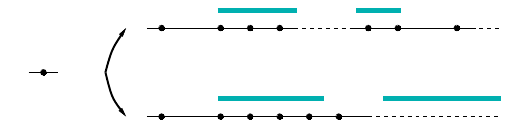}%
\end{picture}%
\setlength{\unitlength}{2072sp}%
\begingroup\makeatletter\ifx\SetFigFont\undefined%
\gdef\SetFigFont#1#2#3#4#5{%
  \reset@font\fontsize{#1}{#2pt}%
  \fontfamily{#3}\fontseries{#4}\fontshape{#5}%
  \selectfont}%
\fi\endgroup%
\begin{picture}(7703,2035)(463,-3725)
\put(2926,-2356){\makebox(0,0)[b]{\smash{{\SetFigFont{6}{7.2}{\rmdefault}{\mddefault}{\updefault}{\color[rgb]{0,0,0}$t$}%
}}}}
\put(7426,-2356){\makebox(0,0)[b]{\smash{{\SetFigFont{6}{7.2}{\rmdefault}{\mddefault}{\updefault}{\color[rgb]{0,0,0}$t'$}%
}}}}
\put(2926,-1861){\makebox(0,0)[b]{\smash{{\SetFigFont{6}{7.2}{\rmdefault}{\mddefault}{\updefault}{\color[rgb]{0,0,0}$\lnot\dabl{G}\phi$}%
}}}}
\put(5446,-1861){\makebox(0,0)[b]{\smash{{\SetFigFont{6}{7.2}{\rmdefault}{\mddefault}{\updefault}{\color[rgb]{0,0,0}$\lnot\dabl{G}\phi$}%
}}}}
\put(7426,-1861){\makebox(0,0)[b]{\smash{{\SetFigFont{6}{7.2}{\rmdefault}{\mddefault}{\updefault}{\color[rgb]{0,0,0}$\evid{G}\phi$}%
}}}}
\put(4726,-2356){\makebox(0,0)[b]{\smash{{\SetFigFont{6}{7.2}{\rmdefault}{\mddefault}{\updefault}{\color[rgb]{0,0,0}$t''$}%
}}}}
\put(2926,-3661){\makebox(0,0)[b]{\smash{{\SetFigFont{6}{7.2}{\rmdefault}{\mddefault}{\updefault}{\color[rgb]{0,0,0}$t$}%
}}}}
\put(2926,-3211){\makebox(0,0)[b]{\smash{{\SetFigFont{6}{7.2}{\rmdefault}{\mddefault}{\updefault}{\color[rgb]{0,0,0}$\lnot\dabl{G}\phi$}%
}}}}
\put(5851,-3211){\makebox(0,0)[b]{\smash{{\SetFigFont{6}{7.2}{\rmdefault}{\mddefault}{\updefault}{\color[rgb]{0,0,0}$\lnot\dabl{G}\phi$}%
}}}}
\put(1126,-2986){\makebox(0,0)[b]{\smash{{\SetFigFont{6}{7.2}{\rmdefault}{\mddefault}{\updefault}{\color[rgb]{0,0,0}$t$}%
}}}}
\put(1126,-2536){\makebox(0,0)[b]{\smash{{\SetFigFont{6}{7.2}{\rmdefault}{\mddefault}{\updefault}{\color[rgb]{0,0,0}$\lnot\dabl{G}\phi$}%
}}}}
\put(586,-2761){\makebox(0,0)[rb]{\smash{{\SetFigFont{6}{7.2}{\rmdefault}{\mddefault}{\updefault}{\color[rgb]{0,0,0}if}%
}}}}
\put(1576,-2671){\makebox(0,0)[lb]{\smash{{\SetFigFont{6}{7.2}{\rmdefault}{\mddefault}{\updefault}{\color[rgb]{0,0,0}then}%
}}}}
\put(1531,-2896){\makebox(0,0)[lb]{\smash{{\SetFigFont{6}{7.2}{\rmdefault}{\mddefault}{\updefault}{\color[rgb]{0,0,0}(or)}%
}}}}
\end{picture}%
\end{center}

\noindent\textbf{C3}~  if $M,t \models_{\logic{time}(\logic{sc})} \dabl{G}\phi$ then (i)~$M,t \models_{\logic{time}(\logic{sc})}
  \evid{G}\phi$, or (ii)~there is $t' < t$, such that $M,t'
  \models_{\logic{time}(\logic{sc})} \evid{G}\phi$ and $M,t''
  \models_{\logic{time}(\logic{sc})} \dabl{G}\phi$ for all $t' < t'' <
  t$.

\begin{center}
  \begin{picture}(0,0)%
\includegraphics{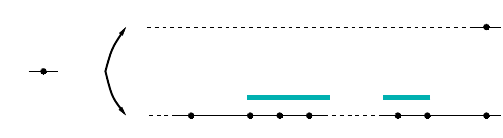}%
\end{picture}%
\setlength{\unitlength}{2072sp}%
\begingroup\makeatletter\ifx\SetFigFont\undefined%
\gdef\SetFigFont#1#2#3#4#5{%
  \reset@font\fontsize{#1}{#2pt}%
  \fontfamily{#3}\fontseries{#4}\fontshape{#5}%
  \selectfont}%
\fi\endgroup%
\begin{picture}(7650,2080)(463,-3770)
\put(5176,-3706){\makebox(0,0)[b]{\smash{{\SetFigFont{6}{7.2}{\rmdefault}{\mddefault}{\updefault}{\color[rgb]{0,0,0}$t''$}%
}}}}
\put(3376,-3211){\makebox(0,0)[b]{\smash{{\SetFigFont{6}{7.2}{\rmdefault}{\mddefault}{\updefault}{\color[rgb]{0,0,0}$\evid{G}\phi$}%
}}}}
\put(5896,-3211){\makebox(0,0)[b]{\smash{{\SetFigFont{6}{7.2}{\rmdefault}{\mddefault}{\updefault}{\color[rgb]{0,0,0}$\dabl{G}\phi$}%
}}}}
\put(7876,-3211){\makebox(0,0)[b]{\smash{{\SetFigFont{6}{7.2}{\rmdefault}{\mddefault}{\updefault}{\color[rgb]{0,0,0}$\dabl{G}\phi$}%
}}}}
\put(7876,-3706){\makebox(0,0)[b]{\smash{{\SetFigFont{6}{7.2}{\rmdefault}{\mddefault}{\updefault}{\color[rgb]{0,0,0}$t$}%
}}}}
\put(3376,-3706){\makebox(0,0)[b]{\smash{{\SetFigFont{6}{7.2}{\rmdefault}{\mddefault}{\updefault}{\color[rgb]{0,0,0}$t'$}%
}}}}
\put(7876,-2311){\makebox(0,0)[b]{\smash{{\SetFigFont{6}{7.2}{\rmdefault}{\mddefault}{\updefault}{\color[rgb]{0,0,0}$t$}%
}}}}
\put(7876,-1861){\makebox(0,0)[b]{\smash{{\SetFigFont{6}{7.2}{\rmdefault}{\mddefault}{\updefault}{\color[rgb]{0,0,0}$\evid{G}\phi$}%
}}}}
\put(1126,-2986){\makebox(0,0)[b]{\smash{{\SetFigFont{6}{7.2}{\rmdefault}{\mddefault}{\updefault}{\color[rgb]{0,0,0}$t$}%
}}}}
\put(1576,-2671){\makebox(0,0)[lb]{\smash{{\SetFigFont{6}{7.2}{\rmdefault}{\mddefault}{\updefault}{\color[rgb]{0,0,0}then}%
}}}}
\put(1126,-2536){\makebox(0,0)[b]{\smash{{\SetFigFont{6}{7.2}{\rmdefault}{\mddefault}{\updefault}{\color[rgb]{0,0,0}$\dabl{G}\phi$}%
}}}}
\put(586,-2761){\makebox(0,0)[rb]{\smash{{\SetFigFont{6}{7.2}{\rmdefault}{\mddefault}{\updefault}{\color[rgb]{0,0,0}if}%
}}}}
\put(1531,-2896){\makebox(0,0)[lb]{\smash{{\SetFigFont{6}{7.2}{\rmdefault}{\mddefault}{\updefault}{\color[rgb]{0,0,0}(or)}%
}}}}
\end{picture}%
\end{center}
\end{definition}

We write $\models_{\logic{lbda}}\phi$ when for every model of deemed ability $M = \tuple{T,<,g}$, and for every instant $t \in T$, we have that $M, t \models_{\logic{time}(\logic{sc})} \phi$.

The logic of ${\logic{lbda}}$ is sound wrt.\ the class of models of deemed ability.
\begin{proposition}
  If $\vdash_{\logic{lbda}}\phi$ then $\models_{\logic{lbda}}\phi$.
\end{proposition}
This is a simple consequence of Proposition~\ref{th:thcompl-timesc} and the fact that the axiom lbda1 (resp.\ lbda2, lbda3) is sound in the class of models of $\logic{time}(\logic{sc})$ with the constraint C1 (resp.\ C2, C3).
Moreover, lbdar1 preserves validity: if $\phi$ is true in every model of $\logic{time}(\logic{sc})$ then it is true in every model of $\logic{lbda}$.

Notice that the constraints correspond to actual conditions on the frames, using the truth value of monolithic formulas (i.e., of formulas in $L_\logic{sc}$) as a shortcut.

To see that lbda1 (resp.\ lbda2, lbda3) is sound in the class of models of $\logic{time}(\logic{sc})$ with the constraint C1  (resp.\ C2, C3), it suffices to spell out the truth condition of the axiom, and see that it corresponds to the constraint.

\para{Interdependence of deemed ability and confirmation}
We say that $\phi$ is \emph{true in a model of deemed ability} $M = \tuple{T,<,g}$, when for every instant $t \in T$, we have that $M, t \models_{\logic{time}(\logic{sc})} \phi$.
To conclude this section, we verify the simple fact that
a deemed ability never occurs in a model of deemed ability
if a confirmation for it never occurs, and the other way around.
Equivalently, a deemed ability occurs at some instant in a model of deemed ability \emph{if and only if} a confirmation for this deemed ability occurs at some (possibly different) instant.
%
\begin{proposition}\label{prop:depend-can-conf}
  $\lnot \dabl{G}\phi$ is true in a model of deemed ability $M$ iff
  $\lnot \evid{G}\phi$ is true in $M$.
\end{proposition}
Suppose $\lnot \dabl{G}\phi$ is true in a model of deemed ability $M = \tuple{T,<,g}$. Hence, $\dabl{G}\phi$ is not true at any instant $t \in T$.
By constraint~1 of Definition~\ref{def:scmodel}, this means that $\evid{G}\phi$ is also not true at any instant, and that $\lnot \evid{G}\phi$ is true in $M$.
Suppose $\lnot \evid{G}\phi$ is true in $M$.
Hence, $\evid{G}\phi$ is not true at any instant $t \in T$.
constraint~C3 of Definition~\ref{def:model-da} makes sure that
$G$ is deemed able to do $\phi$ only if a situation showing confirmation for it has occurred. Since $\evid{G}\phi$ is not true at any instant, $\dabl{G}\phi$ is also not true at any instance, and $\lnot \dabl{G}\phi$ is true in $M$.


\section{Example I: Mele's general practical ability}
\label{sec:elgtemp}
Our core logic of being deemed able serves to explain the existence of
a general ability extending over some period of time, based on
occurrences of confirmations and disconfirmations.
But it does not explain what confirmations and
disconfirmations are. 
It is something that depends on a system designer's choices.
To illustrate the abstract framework, we first
propose to temporalise the logic of
bringing-it-about-that~\cite{Prn77actionsocial,lindahl77}, and more
specifically, Elgesem's extension with agents'
ability~\cite{elgesem97agency,Governatori05Elgesem}. Effectively, we
obtain a temporalised version of Mele's simple
abilities~\cite{mele2003} that are reminiscent of \emph{general
  practical abilities}.

We keep the logic minimal. In~\cite{Tr14jaamas}, more conceivable principles of agency and ability are discussed, and many are rejected. However, any sensible principle (e.g., exploiting the set-theoretical relationships between the acting entities) can find its way into a formalization of a more precise particular domain.

We do not present the model theory. This is a straightforward
extension of the semantics in the models of deemed ability. We present
the extension of the proof theory that we economically signify with
$\vdash$ without index.
That is, in what follows, for any formula $\varphi$, we adopt $\varphi$ as an axiom of an extension of the logic $\logic{lbda}$ by stating $\vdash \varphi$.


\subsection{Bringing-it-about-that}
The \emph{bringing-it-about-that} (BIAT) modality of agency has been used
to model the actions and responsibilities of acting entities. The
formula $E_G\phi$ traditionally reads ``$G$ brings it about that
$\phi$''.

\begin{table}
\begin{center}
\framebox{{\small
 \begin{tabular}{ll}
 {\bf [b1]} & $\vdash \lnot E_G \top$\\
 {\bf [b2]} & $\vdash E_G \phi \limp \phi$\\
 {\bf [b3]} & $\vdash E_G \phi \land E_G \psi \limp E_G (\phi \land \psi)$\\
 {\bf [br1]}& if $\vdash \phi \leqv \psi$ then $\vdash E_G \phi \leqv
 E_G \psi$
\end{tabular}}
}
\end{center}
\caption{Bringing it about that\label{tab:biat}}
\end{table}
The principles of the modality $E_G$ are presented in the
Table~\ref{tab:biat}. By bringing about the truth of a certain proposition, an
acting entity brings about the truth of all equivalent propositions
(rule~br1). Agency in BIAT reflects some responsibility for a
state of affairs. It is stipulated that no acting entity brings about
the tautologies (axiom~b1). Agency requires achievement of results (axiom~b2).
Note that as a
consequence, it is not possible for a group to bring about the
impossible: $\vdash \lnot E_G \bot$.  Actions of acting entities
aggregate (axiom~b3). If at the same instant, an acting entity brings
about two propositions, then it also brings the conjunction of these
propositions. See some details in~\cite{Prn77actionsocial,elgesem97agency,Herzig2018}.





\subsection{Mele's abilities}

Elgesem's logic~\cite{elgesem97agency} is an extension of BIAT with an
elementary notion of ability. The modality $\able{G}\phi$ reads that
``$G$ is able to bring it about that $\phi$''. Elgesem explains at length
that when $G$ brings it about that $\phi$, the state of affairs $\phi$ is
one concerning a
property towards which $G$ has manifested its control. Thus an acting
entity only brings about what it is capable of. It is expressed by the
axiom $E_G\phi \limp \able{G}\phi$. This is exactly what Mele calls
\emph{simple ability}. He writes ``an agent's $A$-ing at a time is
sufficient for his having a simple ability to $A$ at that
time''~\cite[p.~448]{mele2003}. This is distinguished from an
\emph{ability to $A$ intentionally}: ``being able to $A$ intentionally
entails having a simple ability to $A$ and the converse is
false.''~\cite[p.~448]{mele2003}.
A simple ability does not necessarily entail an intentional ability. There can be a simple ability without intention, as doing something, even by accident, is enough ground to have a simple ability.

A major criticism of BIAT is the absence of time, and Elgesem's
extension with abilities is no different. Although we have a means to
infer an ability to $\phi$ from an occurrence of $\phi$-ing, this is of little significance since we cannot reason about what will become of the ability in the
future; Elgesem's system deals only with the present, not the future. Also, ability is only \emph{partially} grounded, because it is consistent that $\lnot
E_G\phi \land \able{G}\phi$ and we cannot reason about past evidence;
again, this system deals only with the present.

However, once integrated in our logic of deemed ability, the problems
are easily overcome by grounding confirmations and disconfirmations. The
resulting notion of ability is similar to what Mele calls \emph{general
  practical ability}. It is an ability which ``we attribute to agents
even though we know they have no opportunity to $A$ at the time of
attribution and we have no specific occasion for their $A$-ing in
mind''~\cite[p.~18]{mele06}.

\subsection{Agency grounded confirmations}

Using the modalities of bringing-it-out,
we are now ready to provide our first axiom of inference of
confirmation of an ability.
\begin{equation*}\label{ax:simple-evid}
  \vdash E_G \phi \limp \evid{G}\phi
  \tag{\textbf{b4}}
\end{equation*}
By axiom~sc1, it follows that $\vdash E_G \phi \limp \dabl{G}\phi$, which
corresponds directly to the principle of simple ability discussed
before. 

Mele says that if an agent $\phi$s, then this is enough to infer that this agent has a simple ability to $\phi$. This is what the principle b4 (with axiom sc1) captures. This is not \emph{necessarily} an intentional ability; for all we know the agent might have $\phi$-ed by accident.

The logic extending $\logic{lbda}$ with the principles adopted
so far is effectively a temporal extension of Elgesem's logic of
agency and ability.

\subsection{Multi-agency ground for confirmations}
We can also exploit the agency of groups of agents more finely.
In a multi-agent setting, axiom~b4 can be
generalised. It is argued in~\cite{Tr14jaamas} that when two groups bring
about some propositions of their own \emph{at the same instant}, they
show that their actions can be carried out together without
conflict.
They might not have worked together at the time, but from $E_{G_1} \phi \land E_{G_2} \psi$ it is plain that the members of the group~$G_1 \cup G_2$ made sure that both $\phi$ and $\psi$ would be true.
We can consider that it is a confirmation that they can together
bring about the conjunction of these propositions. To acknowledge the
superadditive power of groups, we can adopt:
\begin{equation*}\label{ax:m-evid}\vdash E_{G_1} \phi \land E_{G_2} \psi \limp \evid{G_1 \cup G_2} (\phi \land \psi)
  \tag{\textbf{b5}}
\end{equation*}
This subsumes axiom~b4.
Again by axiom~sc1, we obtain the theorem
$\vdash E_{G_1} \phi \land E_{G_2} \psi \limp \dabl{G_1 \cup G_2}
(\phi \land \psi)$ which is an axiom in~\cite{Tr14jaamas}.

A principle analogous to~b5 where the groups' actions are \emph{not} simultaneous would not be acceptable. (This could be represented, although indirectly, by the formula $\dabl{G_1}\phi \land E_{G_2}\psi \limp \evid{G_1 \cup G_2}(\phi \land \psi)$.)
For example, if John brings it about that the light is on at one instant, and Mary brings it about that the light is off one hour later, it is obvious that there is no confirmation of the fact that John and Mary are together able to bring it about that the light is both on and off.

\medskip

Naturally, other principles of confirmations of group ability can be considered depending on the intended application.

In~\cite{Tr14jaamas}, 
to acknowledge the special character of a group without members, any ability of the empty group is rejected. This is stipulated by the formula $\lnot \dabl{\emptyset}\phi$ which is adopted as an axiom.
In our setting, we can adopt the principle $\lnot \evid{\emptyset}\phi$, saying that no situation confirms that the empty group can bring about that $\phi$.
In virtue of Proposition~\ref{prop:depend-can-conf}, it would also imply that $\lnot \dabl{\emptyset}\phi$ as a theorem.
%


%
In presence of groups of agents as sets, principles of monotonicity are natural candidates.
Monotonicity of ability seems even reasonable in some circumstances: if a group~$G$ is deemed able to bring it about that $\phi$, then every group~$G'$ containing $G$ is also deemed able to bring it about that $\phi$. With $G \subseteq G'$, this could be represented by the formula $\dabl{G}\phi \limp \dabl{G'}\phi$. Although a more fundamental principle would be that if a situation confirms that $G$ is able to bring it about that $\phi$ then it also confirms that every group~$G'$ containing $G$ is also deemed able to bring it about that $\phi$. With $G \subseteq G'$, this could be represented by the formula $\evid{G}\phi \limp \evid{G'}\phi$.
%
%

\subsection{Attempt-grounded disconfirmations}
Disconfirmations can also be grounded through the use of agentive attitudes that have been studied in the literature of modal logics for agency.
One can add an abstract notion of attempt to the bringing-it-about-that
framework~\cite{santos97}. As for the $E_G$ modality, we take
$Att_G$ to be a minimal modality:
\begin{equation*} \text{if } \vdash \phi \leqv \psi \text{ then } \vdash Att_G \phi \leqv Att_G \psi
  \tag{\textbf{br2}}
\end{equation*}
Following~\cite{santos97} we could also adopt the axiom (not used
in this paper):
\begin{equation*}
  \vdash E_G\phi \limp Att_G\phi
  \tag{\textbf{b6}}
\end{equation*}
It captures the fact that an action is a sort of attempt; a
successful one by axiom~b2.

Authors such as Kenny~\cite{Kenny75}, have argued that $G$'s ability
to bring about some proposition $\phi$ is $G$'s power to bring about
$\phi$ \emph{when $G$ tries}. We can then have an instance of
disconfirmation of an ability when an acting entity tries to bring about
something but does not actually bring it about. In formula, we have
the following axiom:
\begin{equation*}\label{ax:kenny}
  \vdash Att_G \phi \land \lnot E_G\phi \limp \fals{G}\phi
  \tag{\textbf{b7}}
\end{equation*}

\subsection{A general life cycle of deemed abilities}
Take the logical system extending $\logic{lbda}$ with the principles
of this section. The following deductions can be drawn.
\begin{enumerate}
\item If group~$G$ is not deemed able
to do $\phi$ at some time, $\lnot\dabl{G}\phi$, axiom~lbda2 makes sure
that it is so until some confirmation occurs.
\item Suppose at some later time some acting entities~$G_1, \ldots, G_k$, where $G = G_1 \cup \ldots \cup G_k$, bring about respectively $\phi_1, \ldots \phi_k$ such that $\vdash \phi_1 \land \ldots \land \phi_k \leqv \phi$. By
axiom~b5 and rule~scr2 one can deduce $\evid{G}\phi$.
\item By axiom~sc1 one can deem
$G$ able to bring about $\phi$: $\dabl{G}\phi$.
\item By axiom~lbda1, $G$ will be deemed able to do
$\phi$ until some disconfirmation occurs. 
\item Suppose that at some
later time, $G$ attempts to bring about $\phi$ but does not actually
bring it about, then by axiom~b7 one can infer a
disconfirmation: $\fals{G}\phi$.
\item By axiom~sc2, we infer that $G$ is not deemed able to bring
  about $\phi$: $\lnot\dabl{G}\phi$, and the life cycle is back to step~1.
\end{enumerate}

\section{Example II: Supervised management of deemed abilities}
\label{sec:pract}

In the previous section we have introduced a few notions which allowed us to start grounding confirmations and disconfirmations. We obtained a formalisation of general practical abilities. Leaving the realm of philosophy, we propose further notions to account for when talking about deemed ability. In particular we will consider (i)~supervised management, and (ii)~agreements between a supervisor and an acting entity to accomplish a task before a specified deadline.

Managing deemed abilities in this setting is related to the problem of assessing trust about the ability of an agent or a group of agents to accomplish a task within a multi-agent system (e.g., \cite{10.1007/11755593_3, Wang:2010:ETM:1867713.1867715}).
In this section, we introduce a designated agent that we refer to as the manager of the system. The manager can bring it about that a group~$G$ is deemed able
(within the system of which he is the manager) to bring it about that $\phi$. This can be interpreted as the manager expressing his trust about the ability of the group~$G$ to bring about $\phi$. Similarly, the manager can bring it about that a group~$G$ is \emph{not} deemed able to bring it about that $\phi$, effectively expressing his distrust about the ability of the group~$G$ to bring it about that $\phi$.

\subsection{Supervision-grounded confirmations and disconfirmations}

The logical theory so far leaves the concepts of confirmations and disconfirmations under-specified. This allows a more
flexible management of deemed abilities. Virtually no real system
involving natural agents is a fully autonomous platform. System
managers always reserve themselves means to tweak the system to
reflect managerial decisions that do not always follow the strict
written rules---or logic---of the system.

Let us then assume a special agent~$\mana$ that is the system manager,
or acts on behalf of the system manager.
We are not concerned with the nature of the
manager. It can be a human being, the management team of a MAS-based
business, or a software agent.

In an under-specified system, the manager can make expedient
adjustments to the system with information obtained offline concerning
the qualifications of acting entities. Hence, we will think of ``the
manager brings it about that $G$ is deemed able (resp.\ not deemed able)
to bring about $\phi$'' as an account of confirmation
(resp.\ disconfirmation) in the information system.
\begin{equation*}\vdash E_{\{\mana\}} \dabl{G}\phi \limp \evid{G}\phi
  \tag{\textbf{t1}}
\end{equation*} 
\begin{equation*}\vdash
  E_{\{\mana\}} \lnot \dabl{G}\phi \limp \fals{G}\phi
  \tag{\textbf{t2}}
\end{equation*} 

\para{Scenario} {\em Suppose our rare media service repository is managed 
by $\mana$. If the manager learns that the service $\sigma_3$ has
ceased activity, he can report this information by inserting $E_{\{\mana\}}
\lnot\dabl{G \cup \{\sigma_3\}}\phi$
into the system for every currently existing deemed ability
$\dabl{G \cup \{\sigma_3\}}\phi$, for every group~$G$. The rationale for considering every currently existing deemed ability of every group containing $\sigma_3$ is that a group might be deemed able to bring about something but would need the participation of $\sigma_3$ to do so. With $\sigma_3$ out of business, the manager considers that the group~$G \cup \{\sigma_3\}$ cannot work together,
and this deemed ability should be disconfirmed. Indeed, with axiom~t2 it will
count as a systematic and systemic disconfirmation.

Symmetrically with axiom~t1, the
manager can approve a new piece of specification $\phi$ of a service
$\sigma_4$ by inserting $E_{\{\mana\}} \dabl{\{\sigma_4\}}\phi$ into the system. E.g., suppose $\sigma_4$ is a vinyl record shop. They inform the manager that they received a few original copies of \emph{The Freewheelin' Bob Dylan}; $\phi$ then stands for ``sell \emph{The Freewheelin' Bob Dylan}''.

In general, using both modes of bringing it about that a coalition is deemed able to $\phi$, and bringing it about that a coalition is not deemed able to $\phi$, the manager can adjust the knowledge of the system to any desired change in service specifications.}

\subsection{Task-grounded disconfirmations}

Task negotiation and attribution are other managerial activities that may occur off-platform in an under-specified system concern.
Attribution is worked out by the manager agent, based
supposedly on the information contained in the system. Negotiation can
follow an arbitrary negotiation protocol.

We use $\task{G}(\phi,\psi)$ to denote that the group~$G$ is assigned
the task to bring about the objective $\phi$ before $\psi$ holds. To
serve as a criterion of task identity, we must have:
\begin{equation*}
\text{if } \vdash \phi_1 \leqv \phi_2 \text{ and }  \vdash \psi_1 \leqv \psi_2 \text{ then } \vdash \task{G}(\phi_1,\psi_1) \leqv \task{G}(\phi_2,\psi_2)
\tag{\textbf{tr}}
\end{equation*}
Thus, when the two objectives $\phi_1$ and $\phi_2$ are logically
indistinguishable and so are the two deadlines $\psi_1$ and $\psi_2$,
then the task of $G$ bringing about the objective $\phi_1$ before
$\psi_1$ holds is logically indistinguishable from the task of $G$
bringing about the objective $\phi_2$ before $\psi_2$ holds.

A simple principle then concerns task expiration and completion:
\begin{equation*}\vdash  (\psi \lor E_G\phi) \limp \lnot\task{G}(\phi,\psi)
  \tag{\textbf{t3}}
\end{equation*}
In English, if $\psi$ holds or if $G$ brings it about that $\phi$, then the
group~$G$ is not assigned the task to bring about $\phi$ before $\psi$
holds. The rational is: if $\psi$ holds, then the task has expired,
and if $E_G\phi$ holds then the task has been completed.
By contraposition, whenever $\task{G}(\phi,\psi)$ holds, neither $E_G\phi$ nor $\psi$ do.

\medskip
The manager can negotiate tasks with possible acting entities. We
intend $E_{\{\mana\} \cup G} \task{G}(\phi,\psi)$ to capture the
culmination of a negotiation resulting in the manager and the
group~$G$ agreeing that $G$ will carry the task of bringing about
$\phi$ by the time $\psi$ holds.
For clarity, we use a dedicated vocabulary to express it:
\begin{equation*}\vdash \agree{G}(\phi,\psi) \leqv E_{\{\mana\} \cup G}
\task{G}(\phi,\psi)
\tag{\textbf{t4}}
\end{equation*} Observe that an agreement is
effective, in the sense that $\vdash \agree{G}(\phi,\psi) \limp
\task{G}(\phi,\psi)$ (by axiom~b2).

We can now describe mixed temporal properties of agreements, tasks,
and disconfirmations.

A task originates from a previous agreement and has existed ever since:
\begin{equation*}
  \vdash \task{G}(\phi,\psi) \limp \agree{G}(\phi,\psi) \lor (\task{G}(\phi,\psi) \S \agree{G}(\phi,\psi))
  \tag{\textbf{t5}}
\end{equation*}
A direct reading is that if $G$ is assigned
the task to bring about the objective $\phi$ before $\psi$ holds, then either it has been agreed now, or the task has existed ever since it has been agreed.

An existing task should be maintained at least until it is completed or
its deadline is reached:
\begin{equation*}\vdash \task{G}(\phi,\psi) \limp \task{G}(\phi,\psi) \W (\psi \lor E_G\phi)
  \tag{\textbf{t6}}
\end{equation*}
In fact, together with axiom~t3, it follows that an existing task is maintained exactly until it is completed or the deadline is reached.

\medskip
Finally we can formalise a principle of task-grounded disconfirmation
of ability:
\begin{equation*} 
\vdash (\task{G}(\phi,\psi) \S \agree{G}(\phi,\psi))  \limp
( \psi \land \lnot E_G\phi \limp \fals{G}\phi)
\tag{\textbf{t7}}
\end{equation*}
If $G$ has uninterruptedly had the task to bring about the objective $\phi$ before $\psi$ holds since a time it was agreed, then if the deadline $\phi$ is reached and $G$ still has not completed the task when $\psi$ becomes true, then the situation is a disconfirmation of $G$'s being deemed able to bring about $\phi$ in the system.
The group~$G$ might still be able to bring about $\phi$ eventually. However, having failed to do so under the deadline that was agreed upon with the system's manager, $G$ is not \emph{deemed} able from the point of view of the system.

\para{Scenario} 
{\em Suppose now that the service $\sigma_2$ announces to $\mana$ that it
changed its business model and is able to bring about that $\phi$. We admit that manager
$\mana$ trusts this information and thus brings it about, say at instant
$t$, that $\dabl{\{\sigma_2\}}\phi$. This counts as a confirmation
$\evid{\{\sigma_2\}}\phi$ by axiom~t1. Suppose it is then agreed at instant $t' >
t$ for $\sigma_2$ to do $\phi$ before a reasonable $\psi$:
$\agree{\{\sigma_2\}}(\phi,\psi)$. So by axiom~t4 and axiom~b2, the service $\sigma_2$
is tasked to do $\phi$ before $\psi$. $\task{\{\sigma_2\}}(\phi,\psi)$
holds at $t'$, and will hold by axiom~t6 until the deadline is reached
or $\sigma_2$ brings it about that $\phi$, at which time the task will be
dropped by axiom~t3. Suppose that at the first eventual instant $t'' >
t'$ where $\psi$ holds, it has not yet been the case that
$E_{\{\sigma_2\}}\phi$. So $\fals{\{\sigma_2\}}\phi$ follows from axiom~t7,
and $\lnot\dabl{\{\sigma_2\}}\phi$ follows by axiom~sc2. By axiom~lbda1,
$\dabl{\{\sigma_2\}}\phi$ held between $t'$ and $t''$.}

\section{Conclusions}
The first contribution is a general framework to track, maintain
and reason about the deemed ability of acting entities by taking into
account past and future evidence: confirmations and disconfirmations
(Section~\ref{sec:ebability}). The second contribution is a library of
principles to ground these confirmations and disconfirmations.
These principles permitted us to formalise the philosophically relevant notion of \emph{general practical ability} (Section~\ref{sec:elgtemp}), and to tackle the practically relevant \emph{management of abilities in supervised systems} (Section~\ref{sec:pract}). 

We concentrated on providing a rigorous formal foundation for the
abstract framework and on providing some intuitions on how to put it to use
with some sample instantiations. 

As for now, in the proposed instantiations, deemed ability needs only one
occurrence of actual agency to exist (axiom~b5) and only one failed
attempt (axiom~b7) or failed task (axiom~t7) to disappear. Beyond this first
presentation, more realistic judgements about deemed ability could very easily
be represented; both more sceptical of confirmation and less hasty in accepting disconfirmation. Quickly, however, we will be confronted with the difficulty of deciding how much evidence is enough to justify the confirmation or the disconfirmation of a deemed ability. There might not be a one-size-fits-all solution that is philosophically correct. Nonetheless, a future extension could offer the possibility to parameterise the temporal models with numerical thresholds to fit specific practical domains.

Another extension would allow one to talk about abilities to bring about that a temporal statement, that is, that a temporal formula is true. As for now, the temporalisation and the temporal language $L_{\logic{time}(\logic{sc})}$ does not permit to write, say, 
$\dabl{G}{\F(\phi \land \F \psi)}$ that would capture the fact that the group~$G$ is deemed able to bring about that eventually $\phi$ holds and then that $\psi$ holds at a later time. 
This calls for more expressive temporalisations such as in~\cite{DBLP:journals/ndjfl/FingerG96}.

Formal reasoning about deemed ability is already possible with the
proposed axioms. The various scenarios that illustrate the paper are
examples of it.  The `supposed' events in the scenarios are systemic
events recorded in the execution trace of the repository system. All
the rest is inferred by the logical apparatus to discover new facts in
a way that maintains the system's coherence.

However, our contribution is hardly technical and the formal aspects
are so far admittedly rather straightforward.  The models describe the
trace of a system and are particularly adapted to
simulation. Algorithmic solutions will be a step forward towards using the
framework for the actual tracking and management of abilities in
multi-agent systems.




\section*{Acknowledgements}
This paper has benefited from the comments and suggestions of many readers and reviewers.
I thank them all, as their comments have allowed me to greatly improve this article.
%
I hope that I have responded to these reviews adequately.

\small
\bibliographystyle{plain}
\bibliography{biblio.bib} 

\end{document}